\documentstyle[12pt]{article}

\catcode`\@=11
\long\def\@makefntext#1{
\protect\noindent \hbox to 3.2pt {\hskip-.9pt
$^{{\ninerm\@thefnmark}}$\hfil}#1\hfill}		

\def\@makefnmark{\hbox to 0pt{$^{\@thefnmark}$\hss}}  

\def\ps@myheadings{\let\@mkboth\@gobbletwo
\def\@oddhead{\hbox{}
\rightmark\hfil\ninerm\thepage}
\def\@oddfoot{}\def\@evenhead{\ninerm\thepage\hfil
\leftmark\hbox{}}\def\@evenfoot{}
\def\sectionmark##1{}\def\subsectionmark##1{}}

\newcounter{sectionc}\newcounter{subsectionc}\newcounter{subsubsectionc}
\renewcommand{\section}[1] {\vspace*{0.6cm}\addtocounter{sectionc}{1}
\setcounter{subsectionc}{0}\setcounter{subsubsectionc}{0}\noindent
	{\normalsize\bf\thesectionc. #1}\par\vspace*{0.4cm}}
\renewcommand{\subsection}[1] {\vspace*{0.6cm}\addtocounter{subsectionc}{1}
	\setcounter{subsubsectionc}{0}\noindent
	{\normalsize\it\thesectionc.\thesubsectionc. #1}\par\vspace*{0.4cm}}
\renewcommand{\subsubsection}[1]
{\vspace*{0.6cm}\addtocounter{subsubsectionc}{1}
	\noindent {\normalsize\rm\thesectionc.\thesubsectionc.\thesubsubsectionc.
	#1}\par\vspace*{0.4cm}}

\newcounter{appendixc}
\newcounter{subappendixc}[appendixc]
\newcounter{subsubappendixc}[subappendixc]

\renewcommand{\appendix}[1] {\vspace*{0.6cm}
        \refstepcounter{appendixc}
        \setcounter{figure}{0}
        \setcounter{table}{0}
        \setcounter{equation}{0}
        \renewcommand{\thefigure}{\Alph{appendixc}.\arabic{figure}}
        \renewcommand{\thetable}{\Alph{appendixc}.\arabic{table}}
        \renewcommand{\theappendixc}{\Alph{appendixc}}
        \renewcommand{\theequation}{\Alph{appendixc}.\arabic{equation}}
        \noindent{\bf Appendix \theappendixc #1}\par\vspace*{0.4cm}}

\renewenvironment{thebibliography}[1]
	{\begin{list}{\arabic{enumi}.}
	{\usecounter{enumi}\setlength{\parsep}{0pt}
\setlength{\leftmargin 1.25cm}{\rightmargin 0pt}
	 \setlength{\itemsep}{0pt} \settowidth
	{\labelwidth}{#1.}\sloppy}}{\end{list}}

\newcounter{itemlistc}
\newcounter{romanlistc}
\newcounter{alphlistc}
\newcounter{arabiclistc}

\newcommand{\fcaption}[1]{
        \refstepcounter{figure}
        \setbox\@tempboxa = \hbox{\footnotesize Fig.~\thefigure. #1}
        \ifdim \wd\@tempboxa > 6in
           {\begin{center}
        \parbox{6in}{\footnotesize\baselineskip=12pt Fig.~\thefigure. #1}
            \end{center}}
        \else
             {\begin{center}
             {\footnotesize Fig.~\thefigure. #1}
              \end{center}}
        \fi}

\newcommand{\tcaption}[1]{
        \refstepcounter{table}
        \setbox\@tempboxa = \hbox{\footnotesize Table~\thetable. #1}
        \ifdim \wd\@tempboxa > 6in
           {\begin{center}
        \parbox{6in}{\footnotesize\baselineskip=12pt Table~\thetable. #1}
            \end{center}}
        \else
             {\begin{center}
             {\footnotesize Table~\thetable. #1}
              \end{center}}
        \fi}

\def\@citex[#1]#2{\if@filesw\immediate\write\@auxout
	{\string\citation{#2}}\fi
\def\@citea{}\@cite{\@for\@citeb:=#2\do
	{\@citea\def\@citea{,}\@ifundefined
	{b@\@citeb}{{\bf ?}\@warning
	{Citation `\@citeb' on page \thepage \space undefined}}
	{\csname b@\@citeb\endcsname}}}{#1}}

\newif\if@cghi
\def\cite{\@cghitrue\@ifnextchar [{\@tempswatrue
	\@citex}{\@tempswafalse\@citex[]}}
\def\citelow{\@cghifalse\@ifnextchar [{\@tempswatrue
	\@citex}{\@tempswafalse\@citex[]}}
\def\@cite#1#2{{$\null^{#1}$\if@tempswa\typeout
	{IJCGA warning: optional citation argument
	ignored: `#2'} \fi}}

 1
 1
 1

\font\ninerm=cmr9

\begin{document}
\noindent\hspace*{\fill}{\Large \bf Chiral Restoration in the Early Universe:
 }\hspace*{\fill}\\
\noindent\hspace*{\fill}{\Large \bf Pion Halo in the Sky     }\hspace*{\fill}\\

\noindent\hspace*{\fill}{  Ngee-Pong Chang ({ \em npccc@cunyvm.cuny.edu\ })
}\hspace*{\fill}\\
\noindent\hspace*{\fill}{  Department of Physics}\hspace*{\fill}\\
\noindent\hspace*{\fill}{  City College \& The Graduate School of City
University of New York}\hspace*{\fill}\\
\noindent\hspace*{\fill}{  New York, N.Y. 10031 }\hspace*{\fill}\\

\vspace*{-.1in}
\noindent\hspace*{\fill}\parbox{4.5in}{\em
\noindent
           The thermal vacuum at high $T$ is chiral invariant, but
           is it the same chirality we knew at $T=0$ ?
           New class of order parameters is proposed to probe this.
	   The pion remains a Nambu-Goldstone and propagates with a halo.
}\hspace*{\fill}$\;$

\newcommand{\gn}{\mbox{$\gamma_{\stackrel{}{5}}$}}
\newcommand{\adag}{a^{\dagger}_{p,s}}
\newcommand{\atildedag}{\tilde{a}^{\dagger}_{-p,s}}
\newcommand{\bdag}{b^{\dagger}_{-p,s}}
\newcommand{\btildedag}{\tilde{b}^{\dagger}_{-p,s}}
\newcommand{\apsbeta}{a^{\beta}_{p,s}}
\newcommand{\apsbetadag}{a_{-p,s}^{\beta\dagger}}
\newcommand{\adagbdag}{a^{\dagger}_{p,s} b^{\dagger}_{-p,s}}
\newcommand{\aps}{a^{}_{p,s}}
\newcommand{\bps}{b^{}_{-p,s}}
\newcommand{\bpsbeta}{b^{\beta}_{p,s}}
\newcommand{\bpsbetadag}{b_{-p,s}^{\beta\dagger}}
\newcommand{\Adag}{A^{\dagger}_{p,s}}
\newcommand{\Bdag}{B^{\dagger}_{-p,s}}
\newcommand{\Aps}{A^{}_{p,s}}
\newcommand{\Apsbeta}{A^{\beta}_{p,s}}
\newcommand{\Apsbetadag}{A^{\beta\dagger}_{-p,s}}
\newcommand{\Bps}{B^{}_{p,s}}
\newcommand{\Bpsbeta}{B^{\beta}_{p,s}}
\newcommand{\Bpsbetadag}{B^{\beta\dagger}_{-p,s}}
\newcommand{\ApL}{A^{}_{p,L}}
\newcommand{\BpL}{B^{}_{-p,L}}
\newcommand{\ApR}{A^{}_{p,R}}
\newcommand{\BpR}{B^{}_{-p,R}}
\newcommand{\apL}{a^{}_{p,L}}
\newcommand{\bpL}{b^{}_{-p,L}}
\newcommand{\apR}{a^{}_{p,R}}
\newcommand{\bpR}{b^{}_{-p,R}}
\newcommand{\AdagL}{A^{\dagger}_{p,L}}
\newcommand{\AdagR}{A^{\dagger}_{p,R}}
\newcommand{\BdagL}{B^{\dagger}_{-p,L}}
\newcommand{\BdagR}{B^{\dagger}_{-p,R}}
\newcommand{\adagL}{a^{\dagger}_{p,L}}
\newcommand{\adagR}{a^{\dagger}_{p,R}}
\newcommand{\bdagL}{b^{\dagger}_{-p,L}}
\newcommand{\bdagR}{b^{\dagger}_{-p,R}}
\newcommand{\eps}{\epsilon}
\newcommand{\gnplus}{\gamma \cdot n_{_{+}}}
\newcommand{\gnminus}{\gamma \cdot n_{_{-}}}
\newcommand{\gnplusdef}{\left( \vec{\gamma} \cdot \hat{n}-\gamma_o \right)}
\newcommand{\gnminusdef}{\left( \vec{\gamma} \cdot \hat{n}+\gamma_o \right)}
\newcommand{\abab}{a^{\dagger}_{p,L}\,b^{\dagger}_{-p,L}\,a^{\dagger}_{p,R}
                   \,b^{\dagger}_{-p,R}}
\newcommand{\alphai}{\alpha_{i}}
\newcommand{\limit}{\lim_{\Lambda^2 \rightarrow \infty}}
\newcommand{\p}{\vec{p}, p_o}
\newcommand{\poprime}{p_o^{\prime}}
\newcommand{\prodps}{\prod_{p,s}}
\newcommand{\prodp}{\prod_{p}}
\newcommand{\psibar}{\bar{\psi}}
\newcommand{\psibarpsi}{ < \bar{\psi} \, \psi
            > }
\newcommand{\PsibarPsi}{ < \bar{\Psi} \, \Psi
            > }
\newcommand{\psibeta}{\psi^{}_{\beta}}
\newcommand{\psibarbeta}{\bar{\psi}_{\beta}}
\newcommand{\psidag}{\psi^{\dagger}}
\newcommand{\psidagbeta}{\psi^{\dagger}_{\beta}}
\newcommand{\psiL}{\psi_{_{L}}}
\newcommand{\psiR}{\psi_{_{R}}}
\newcommand{\Q}{Q_{_{5}}}
\newcommand{\Qa}{Q_{_{5}}^{a}}
\newcommand{\Qbeta}{Q_{5}^{\beta}}
\newcommand{\qqbar}{q\bar{q}}
\newcommand{\sumps}{\sum_{p,s}}
\newcommand{\thetap}{\theta_{p}}
\newcommand{\costhetap}{\cos{\thetap}}
\newcommand{\sinthetap}{\sin{\thetap}}
\newcommand{\thetaset}{\{ \thetap  \}}
\newcommand{\thetapi}{\thetap{}_{i}}
\newcommand{\Tomega}{\frac{\Tprime}{\omega}}
\newcommand{\pomega}{\frac{p}{\omega}}
\newcommand{\Tprime}{T'}
\newcommand{\Tprimesq}{T^{'2}}
\newcommand{\vac}{| vac \rangle}
\newcommand{\vacbeta}{| vac \rangle_{_{\beta}}}
\newcommand{\x}{\vec{x},t}
\newcommand{\xPrime}{\vec{x} - \hat{n} (t - t' ), t'}
\newcommand{\xPrimet}{\vec{x} + \hat{n} (t - t'), t'}
\newcommand{\y}{\vec{y}, y_o}

\section{Introduction}
	The vanishing of $\psibarpsi$ has been often cited as the reason
        for the evidence for the restoration of chiral symmetry.  But is
        the chiral symmetry at high temperatures the same chirality
	that we know at $T=0$?
	My investigations have shown that the `restored'
	chirality is an interesting {\em  morphosis} of the old zero temperature
	chirality.

	The original NJL vacuum undergoes a
	{\em  new phase transformation} so that the new generalized
	NJL vacuum has a $90^{o}$ phase. This $90^{o}$ is responsible
	for $\psibarpsi$ vanishing at high $T$.
	By a study of the spacetime quantization of the effective action
	of a fermion propagating through a hot environ, I have found that
	the factor of $i$ in the thermal vacuum may be traced to
 	the presence of a spacelike cut in the hot fermion propagator.

	The new vacuum continues to break our zero temperature chirality.
	The pion thus remains a Nambu-Goldstone boson,
	and actually acquires a halo while propagating through the early
	universe.

	The pion is a messenger of an underlying broken symmetry of the
	universe, {\em  viz.} that of chirality, under the transformation
$
	\psi (\x)  \rightarrow  {\rm e}^{i \alpha \gn} \; \psi (\x).
$
	The chiral charge, $\Q$, which generates this transformation
$
	\Q = \int d^3 x  \;\psi^{\dagger} (\x)
			\gn \psi (\x)
$
	does not annihilate the vacuum.  Instead, acting on the NJL
	vacuum, it generates, up to a normalization factor, the
	state for a {\em  zero momentum pion},
$
	 \;\Q |vac> \;\propto\;  | \vec{\pi} (\vec{p} = 0 ) \rangle
$
	where ($s= \mp 1$ for $L,R$ helicities)
\begin{equation}
	| vac > \;=\; \prod_{p,s} \left( \costhetap
	\; \;-\; \, s \,  \sinthetap \, \adagbdag
			\right) \; | 0 >		\label{eq-NJL-vac}
\end{equation}
	The angle, $\thetap$, is related to the mass $m$
	acquired by the fermion through the NJL gap equation
$
   	\tan{2\thetap}  =  \frac{m}{p}   	     	
$

	Using the fact that $\Q$ is a constant of motion, it is easy to
	show directly that this zero momentum pion, $\Q |vac>$,
	has zero energy, thus confirming the status of the pion as a QCD
	Nambu-Goldstone boson.

        A signature of this dynamical symmetry breaking is the familiar order
	parameter, $\psibarpsi$.
	For $T > T_c$, however, it is well known that
	$\psibarpsi$ vanishes.  Chiral symmetry is said to be restored
	at $T_c$, but is it the {\em  same old chiral symmetry we knew
	at $T=0$ ?}

\section{High Temperature Effective Action}
	At high temperatures, lattice work as well as continuum field theory
	calculations show that the effective action indeed exhibits a manifest
	chiral symmetry. In thermal field theory, there is
	the famous (BPFTW) Braaten-Pisarski Frenkel-Taylor-Wong
	action\cite{BP}
	that describes the
	propagation of a QCD fermion through a hot medium
	($T^{'2}  \equiv \frac{\textstyle g_r^2 }{\textstyle 3} T^2$,
	while the angular brackets denote an average over the orientation
	$\hat{n}$)
\begin{equation}
   {\cal L}_{\rm eff} = - \psibar \gamma_{\mu} \partial^{\mu}
                          \psi
                       - \frac{T^{'2}}{2\;\;} \, \psibar
                         \left<
                       \frac{\gamma_o - \vec{\gamma} \cdot \hat{n} }
                       {D_o + \hat{n} \cdot \vec{D} }
                         \right> \psi          		\label{eq-BP-action}
\end{equation}
	and we see the global chiral symmetry of the action.  But the
	{\em  nonlocality} of the action implies that the Noether charge
	for this new chirality is not the same as the $T=0$ chirality.

	The fermion propagator\cite{Weldon-Klimov} that results from
	this action shows a pseudo-Lorentz invariant particle pole of
	mass $\Tprime$ (the so-called thermal
	mass).  But, in addition,
	there is a pair of conjugate {\em  spacelike} plasmon cuts in the
	$p_o$-plane that run just above and below the real
	axis\cite{Chang-xc}, from $p_o = -p$ to $p_o = p$.  The cuts
	are associated with the logarithms that result from the angular
	average in eq.(\ref{eq-BP-action}).  Along the real $p_o$
	axis, the propagator function has been chosen real. As a result,
	for $t>0$, say, the propagator function takes the form
\begin{eqnarray}
                < \psi(x) \bar{\psi} (0) >
	&=&	\;\;\; \int \frac{d^3 p}{ (2\pi)^3 } \;
		{\rm e}^{i \vec{p} \cdot \vec{x}} \;\left\{
		 Z_{p} \frac{-i \vec{\gamma}
			\cdot \vec{p} + i \gamma_o \omega }{2 \omega} \;
		{\rm e}^{ - i \omega t}   \right. \nonumber\\
	& &	-  \frac{\Tprimesq}{8\;\;} \;
		\left.\int_{-p}^{p} \,\frac{dp_o'}{p^3}  \;
		\frac{i \vec{\gamma} \cdot \vec{p} p_o'
		- i \gamma_o p^2}{p^2 - p_o^2 + \Tprimesq}   \;
		{\rm e}^{- i p_o' t}
                \;+\; O (T'^4)
		\right\}			\label{eq-spacelike-cut}
\end{eqnarray}
	Note that the spinor structure of the massive particle pole term
	has the (unusual) feature of being manifestly chiral invariant.
	The wave function renormalization constant $Z_p$ at the pole
	is given by
$
	Z_{p} = 1 - \frac{\Tprimesq}{4p^2} \left( \ln{ \frac{4p^2}{\Tprimesq}}
					- 1 \right)
$

	In a recent study of the spacetime quantization of the
	BPFTW action\cite{Chang-bp-local}, I have shown that
	the spacelike cuts dictate a new thermal vacuum of the
	type
\begin{equation}
	| vac' > \;=\; \prod_{p,s} \left( \costhetap
		 \;-\; i\, s \, \sinthetap \, \adagbdag
			\right) \; | 0 >	\label{eq-new-vac}
\end{equation}
	The $90^{o}$ phase here in the generalized NJL vacuum is
	the reason why $\psibarpsi$ vanishes for $T \geq
	T_c$.

	The quantization of a nonlocal action is a very technical
	matter.  Suffice it here to say that the quantization has been
	formulated in terms of auxiliary fields so that the resulting action
	is local. In this context, the pseudo-Lorentz particle pole is
	described in terms of the massive canonical Dirac field, $\Psi$,
	and the spacelike cuts are associated with the auxiliary fields, which
	are functions of $\Psi$.
	This formulation allows for a systematic expansion
	of the $\psi$ field in terms of the massive canonical Dirac field,
	$\Psi$.
	Let the $t=0$ expansion for the original massless $\psi$ field read
\begin{equation}
	\psi (\vec{x}, 0) = \frac{1}{\sqrt{V}}  \sum_{p}
		\; {\rm e}^{i \vec{p} \cdot \vec{x}}
		\left\{ \left( \begin{array}{r}
			\chi_{_{p,L}} a^{}_{p,L}  \\
			\chi_{_{p,R}} a^{}_{p,R} \\
			\end{array}\right)  \;+\;
                \left( \begin{array}{r}
			\chi_{_{p,R}} b^{\dagger}_{-p,R} \\
			\;-\;
			\chi_{_{p,L}} b^{\dagger}_{-p,L}
			\end{array} \right) \right\}
\end{equation}
	with a corresponding canonical expansion for the massive $\Psi$,
	then we find
\begin{eqnarray}
	\aps	&=& \Aps \;-\; i \;s\; \frac{\Tprime}{2 p}
		    \Bdag + O(\Tprime {}^2 )	\label{eq-aps-Aps-1} \\
	b^{}_{p,s} &=& B^{}_{p,s} \;+\; i \;s\; \frac{\Tprime}{2 p}
		    A^{\dagger}_{-p,s}
			+ O(\Tprime {}^2)	\label{eq-bps-Bps-1}
\end{eqnarray}
	The $O(\Tprime)$ terms in the Bogoliubov transformation
	imply the new thermal vacuum of eq.(\ref{eq-new-vac}).

	The chiral charge at high $T$ is given by
\begin{equation}
        Q_{5}^{\beta} =  - \frac{1}{2} \; \sum_{p,s}\; s \;
                      \left(
                      A^{\dagger}_{p,s} A^{}_{p,s} + B^{\dagger}_{-p,s}
                      B^{}_{p,s}
                                  \right)
\end{equation}
	so that it annihilates the new thermal vacuum,
        in direct contrast with the $T=0$ Noether charge
\begin{equation}
        Q_{_{5}}
        = - \frac{1}{2}
             \sum_{p,s}\, s \; \left( a^{\dagger}_{p,s} a^{}_{p,s} +
             b^{\dagger}_{-p,s} b^{}_{-p,s}
			\right)			\label{Q5}
\end{equation}
	which clearly fails to annihilate the vacuum at high $T$.

	Our results on the space-like cut of the thermal propagator
	suggests a new class of order parameters that probe the physics
	of this space-like cut.  The usual order parameters involve
	correlators of the local bilinear $\psibarpsi$ and probe the
	time-like spectrum.  To probe the space-like support of the
	spectrum, we would have to consider the new class of operators
\begin{eqnarray}
	< vac |  \; \int_{-\infty}^{\infty} dt'
		\int \frac{d^3 \hat{n}{}^{\prime}}{4\pi}
         \bar{\psi}(\vec{x}, t) \;\hat{n}{}^{\prime} \cdot \vec{\nabla}
                \;\psi(\vec{x} - \hat{n}{}^{\prime} t', t') \;|vac>
\end{eqnarray}
	which, by construction, probe the space-like spectrum of the
	thermal operator.

\section{Pion halo in the Sky}

	The pion we know at zero temperature is not massless,
	but has a mass of $135 \;MeV$.
	At very high $T$, when electroweak symmetry is restored, we
	have the interesting new possibility that the pion will fully
	manifest its Nambu-Goldstone nature and remain physically
	massless.\cite{Chang-QCD}

	The pion is described by an interpolating field operator,
$
	 \pi^{a}  \;\sim\;   i \bar{\psi} \gn T^{a} \; \psi
$
	which does not know about temperature. It is the vacuum that
	depends on $T$.
	The state vector for a zero momentum pion at high $T$ may
	be obtained from the thermal vacuum by the action
$
	\Q^{a} | vac' > \;\propto \; | \pi^{a} ( \vec{p} = 0 ) \rangle
$
	This pion now has the property that even though it is
	massless, it can acquire a {\em  screening mass} proportional
	to $T$.
	This is the pion mass that has
	been measured on the lattice at high $T$.

	As a result, the pion propagates in the early universe with a
	halo.  The retarded function for the pion shows that the
	signal propagates along the light cone, with an additional
	exponentially damped component coming from the past history
	of the source.
\begin{equation}
        D_{\rm ret} (\x) = \theta (-t) \left\{ \delta(t^2 - r^2)
			+ \frac{\Tprime}{r} \theta(t^2 - r^2)
                    \left[ {\rm e}^{-\Tprime | t-r| }
                    	+  {\rm e}^{-\Tprime | t+r| }
				\right] \right\}
\end{equation}
        The screening mass leads to an accompanying modulator
        signal that `hugs' the light cone, with a screening length
        $\propto 1/T$.

	What are the cosmological consequences of such a pion in the
	alphabet soup of the early universe?

        The Nambu-Goldstone theorem forces the pion to remain a strictly
        massless bound state at high $T$, so that the pion does not
        dissociate but continues to contribute to the partition function
        of the early universe.
	Fortunately, the pion does not contribute so many degrees of freedom
	as to upset the usual picture of the cooling of the universe.  But
	I leave it to experts to help figure out the subtle changes there
	must surely be in the phase transitions of the early universe.

	In the beginning there was light, and quarks, and gluons, to which
	we must now add the pions with halo.

\end{document}